\documentclass{jaa}
\usepackage{natbib}
\usepackage{graphicx}	
\usepackage{amsmath}	
\usepackage{amssymb}	
\usepackage{bm}		
\usepackage{subfigure}
\usepackage{hyperref}
\usepackage{booktabs}

\newcommand{\kms}{\,km\,s$^{-1}$} 
\newcommand{\tco}{$^{13}$CO(3-2)}
\newcommand{\cetno}{C$^{18}$O(3-2)}
\newcommand{\co}{CO(3-2)}
\newcommand{\comment}[1]{}
\newcommand{\arcsec}{''}
\newcommand{\bdstar}{BD\,+45\,3216}
\newcommand{\rmsHII}{G083.7962+03.3058}
\newcommand{\rmsYSO}{G083.7071+03.2817}

\begin{document}\sloppy

\title{Investigating the morphology and CO gas kinematics of Sh2-112 region}


\author{Kshitiz K. Mallick\textsuperscript{1},
        Saurabh Sharma\textsuperscript{1},
        Lokesh K. Dewangan\textsuperscript{2},
        Devendra K. Ojha\textsuperscript{3},
        Neelam Panwar\textsuperscript{1},
        Tapas Baug\textsuperscript{4}}
\affilOne{\textsuperscript{1}Aryabhatta Research Institute of Observational Sciences (ARIES),
                             Manora Peak, Nainital, 263002, India.\\}
\affilTwo{\textsuperscript{2}Physical Research Laboratory, Navrangpura, Ahmedabad 380009, India.\\}
\affilThree{\textsuperscript{3}Department of Astronomy and Astrophysics,
             Tata Institute of Fundamental Research, Homi Bhabha Road,
             Mumbai 400005, India\\}
\affilFour{\textsuperscript{4}S.N. Bose National Centre for Basic Sciences, Block JD,
             Sector III, Salt Lake, Kolkata 700106, West Bengal, India}



\twocolumn[{

\maketitle
\corres{kshitiz@aries.res.in}
\msinfo{1 January 2015}{1 January 2015}

\begin{abstract}
We present a study of the molecular cloud in Sh2-112 massive star forming
region using the 3--2 transition of CO isotopologues - CO, $^{13}$CO,
and C$^{18}$O; supplemented in part by CGPS H\,I line emission and MSX
data.
Sh2-112 is an optically visible region powered by an O8V type massive star
\bdstar, and hosts two Red MSX Survey sources -- \rmsHII\, and \rmsYSO\, --
classified as H\,II region and young stellar object, respectively.
Reduced spectral data products from the James Clerk Maxwell Telescope
archive, centered on the two RMS objects with $\sim$7'$\times$7' field
of view each, were utilised for the purpose.
The \tco\, channel map of the region shows the molecular cloud to have
filamentary extensions directed away from the massive star, which also
seems to
be at the edge of
a cavity like structure.
Multiple molecular cloud protrusions into this cavity structure host
local peaks of emission.
The integrated emission map of the region constructed from only
those emission clumps detected above 5$\sigma$ level in the
position-position-velocity space affirms the same.
MSX sources were found distributed along the cavity boundary
where the gas has the been compressed. Spectral extraction at these positions
yielded high Mach numbers and low ratios of thermal to non-thermal pressure,
suggesting a dominance of supersonic and non-thermal motion in the cloud.

\end{abstract}


}]



\doinum{12.3456/s78910-011-012-3}
\artcitid{\#\#\#\#}
\volnum{000}
\year{0000}
\pgrange{1--}
\setcounter{page}{1}
\lp{1}


\section{Introduction}

Formation and evolution of massive stars is a significant area of research
\citep{Zinnecker_2007ARAA}, as such stars can affect the evolution of the
Galaxy via their immense matter and radiation output. Though it has been
difficult to establish an evolutionary sequence for massive stars, due to
their faster evolution and relatively rarer occurrences, various high-mass
precursors have been sought to be established, such as massive cold molecular
cores, massive starless clumps, infrared dark clouds (IRDCs)
\citep{Motte_2018ARAA}, massive young stellar objects (MYSOs)
\citep{Hoare_2005IAUS}, and so on.
The Lyman continuum radiation from high-mass stars
ionizes the surrounding medium, and the subsequent expanding H\,II regions
have been subjects of investigation for understanding the
impact of radiation output in triggering or quenching further star formation
in the natal cloud \citep{Elmegreen_1998ASPC,Elmegreen_2011EAS,Ogura_2010ASInC}.

\begin{figure}
\includegraphics[trim={0.5cm 1.cm 2.cm 2cm},clip,width=\linewidth]{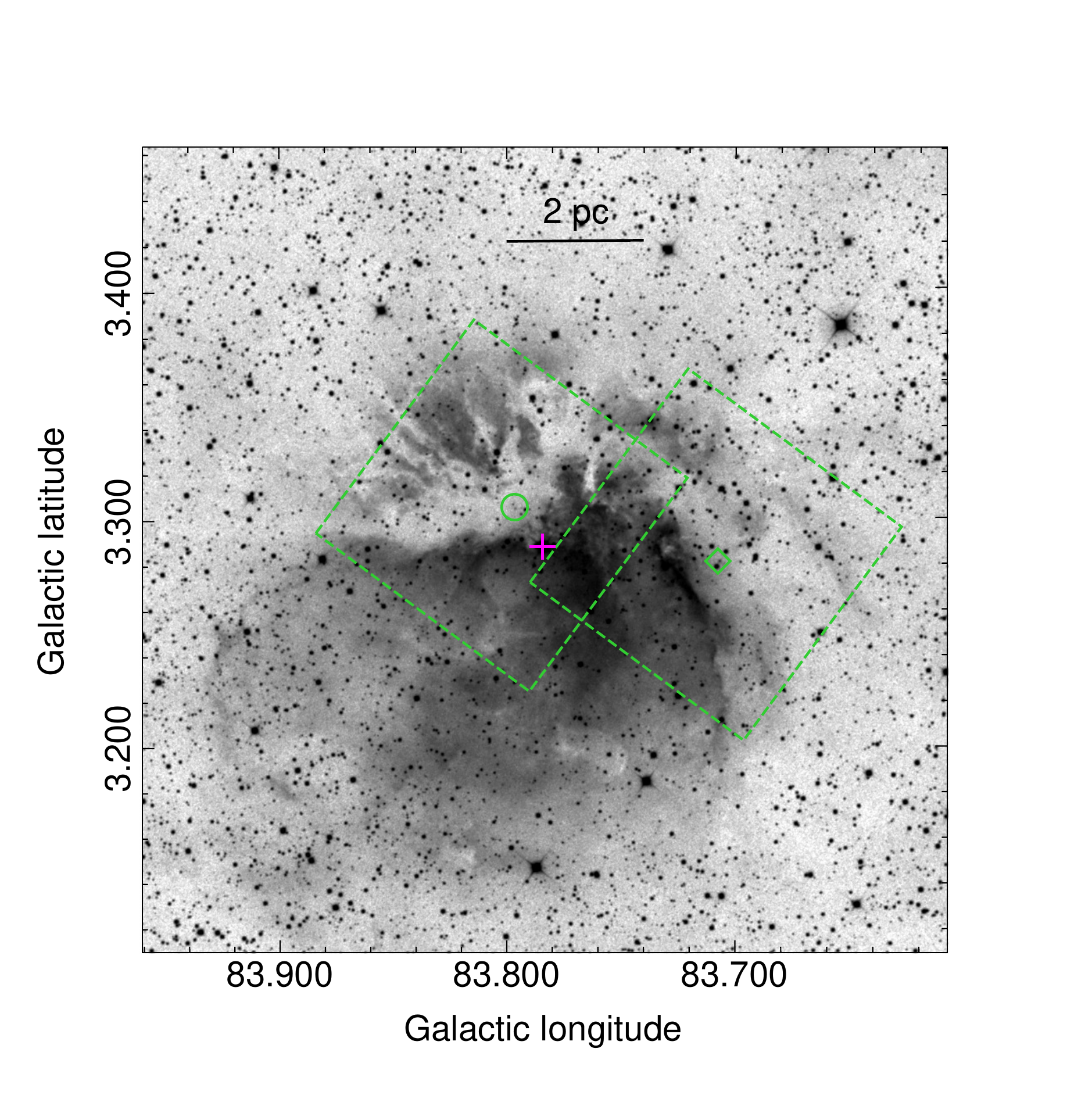}
\caption{DSS2-Red image of the
Sh2-112 region. Plus symbol marks the massive
star \bdstar. Circle and diamond denote the locations of the two RMS
sources, \rmsHII\, and \rmsYSO, respectively. Dashed boxes show the
field of view of the two JCMT fields.}
\label{fig_DSS}
\end{figure}

Sh2-112 is an optically visible H\,II region \citep{Dickel_1969AA}
(Figure \ref{fig_DSS})
(\emph{l}\,$\sim$\,83.758\textdegree, \emph{b}\,$\sim$\,+03.275\textdegree)
at a distance of 2\,kpc \citep{Panwar_2020} in the northern Galactic plane.
It is associated with the Cygnus superbubble, and is one of the many Sharpless
\citep{Sharpless_1959} H\,II regions in Cygnus \citep{Uyaniker_2001AA}.
The region appears nearly circular at optical wavelengths, and displays
a blister morphology in radio emission \citep{Israel_1978}.
Figure \ref{fig_DSS} displays a prominent dark lane against the optical
emission. The massive star \bdstar\, -- which has been estimated to be of
spectral type O8V \citep{Lahulla_1985,Panwar_2020} -- lies close to this lane,
offcenter from the optical nebulosity.
The Red MSX Source (RMS) survey by \citet{Lumsden2013_RMSPaper}
-- which aims to
catalog massive young stellar population in the Galaxy -- has identified
two sources associated with the Sh2-112 region, namely \rmsHII\, and \rmsYSO,
classified as H\,II region and YSO, respectively, in the catalog.
Though the molecular cloud in this region has been subject to investigations
in literature in various molecular transitions, such as
HCN (J=1-0) \citep{Burov_1988PAZh}, CO \citep{Blitz_1982ApJS},
$^{13}$CO (J=1-0) \citep{Dobashi_1994ApJS,Dobashi_1996ApJ}, and
CO isotopologues
\citep{Urquhart_13CO_RMS_2008AA,Maud_2015MNRAS_Outflows,Panja_2022},
they have either been at a low resolution, and/or as a
part of a large statistical study or a larger region encompassing Sh2-112.
As such, a detailed examination of the molecular cloud associated with
Sh2-112 H\,II region is pending, a void we try to fill in this paper.

The organisation of this paper is as follows. In section \ref{section_dataused},
we list the datasets used and any processing steps. This is followed by
section \ref{section_results}, where we present the analysis results for the
molecular gas kinematics. Finally we discuss our results in
section \ref{section_discussion}, followed by a summary and conclusions
section in section \ref{section_summary}.

\section{Data Used}
\label{section_dataused}

\subsection{Archival Spectral Data Products}

We obtained the archival reduced and calibrated JCMT
(\emph{James Clerk Maxwell Telescope}) spectral cubes from the
Canadian Astronomy Data
Centre\footnote{https://www.cadc-ccda.hia-iha.nrc-cnrc.gc.ca/en/} (CADC),
for
\co\, (rest frequency = 345.79599 GHz),
\tco\, (rest frequency = 330.587960 GHz),and
\cetno\, (rest frequency = 329.330545 GHz).
The \emph{J=3--2} transition traces gas at a critical density
of $\sim$\,10$^{4-5}$\,cm$^{-3}$ \citep{Buckle_2010MNRAS}.
In the direction of the Sh2-112 region, two fields have been observed by
the JCMT, namely G083.7962+03.3058 and G083.7071+03.2817
(see Figure \ref{fig_DSS}), using the
HARP/ACSIS
\citep[Heterodyne Array Receiver Programme/Auto-Correlation
Spectral Imaging System;][]{Buckle_HARP_ACSIS_2009}
spectral imaging system. Cubes for both of these fields were retrieved.
The temperature scale used for the pixel brightness units is
T$_A^*$ (antenna temperature).
A basic processing using the
\textsc{starlink kappa} \citep{Currie_starlink_2014}
package was carried out, wherein the downloaded reduced cubes for the
two fields were mosaiced; the spectral axis was converted to LSRK
velocity scale; and the coordinate system was transformed to Galactic from
FK5 for ease of analysis.
Thereafter the cubes were rebinned along spectral axis to 0.5\kms\,
channel width.
Table \ref{table_data_jcmt} provides the details of the JCMT data used.
While the rebinned cubes were used for the detection of spatial structures
(sections \ref{section_chmaps} and \ref{section_momentmaps})
as they have lower RMS noise; for the calculation of physical parameters
(section \ref{section_parameters})
the native channel width ($^{13}$CO and/or C$^{18}$O) cubes were used as
high velocity resolution is required for the same.

For the Sh2-112 region, we also procured archival spectral cube for
21\,cm H\,I line emission produced by the Canadian Galactic Plane
Survey (CGPS) Consortium \citep{Taylor_CGPS_2003AJ}.
We use the cube on an as is basis for examining the morphology of the
region. The 21\,cm cube has an angular resolution of
$\sim$1'$\times$1'\,cosec$\delta$, channel width of $\sim$\,0.8\kms,
and a velocity resolution of $\sim$\,1.3\kms.

\begin{table*}
\centering
\caption{JCMT Data Used}
\label{table_data_jcmt}
\begin{tabular}{lllll}
\hline
Line/Wavelength   &  Spatial     & Native Channel & Rebinned Channel & Program  \\
                  &  Resolution  & Width          & Width (\& Noise) & IDs Used \\
\hline
\co    &  $\sim$14\arcsec  & $\sim$0.42\kms  & 0.5\kms & M08AU19 \\
       &                   &                 & ($\sim$0.37\,K) & \\
\cmidrule(lr){2-5}
\tco   &  $\sim$14\arcsec  & $\sim$0.05\kms  & 0.5\kms & M08AU19, \\
       &                   &                 & ($\sim$0.35\,K) & M08BU18 \\
\cmidrule(lr){2-5}
\cetno &  $\sim$14\arcsec  & $\sim$0.05\kms  & 0.5\kms & M08AU19, \\
       &                   &                 & ($\sim$0.45\,K) & M08BU18 \\
\hline
\end{tabular}
\end{table*}

\subsection{Archival Imaging Data Products}

We obtained the stellar sources present in MSX (Midcourse Space Experiment)
Catalog 6 \citep{Egan_2003} for our JCMT field of view, via the NASA/IPAC
Infrared Science
Archive\footnote{https://irsa.ipac.caltech.edu/}. Though observations
by the MSX satellite were carried out in 6 bands ranging from 4-21\,$\mu$m,
the most sensitive observation has been in the mid-infrared (MIR)
A band (8.28\,$\mu$m) \citep{Lumsden_MSX_2002}.
The other three MIR bands used by
MSX are C (12.13\,$\mu$m), D (14.65\,$\mu$m), and E (21.3\,$\mu$m).
MSX has surveyed the Galactic plane at MIR bands with a spatial resolution
of $\sim$\,18.3\arcsec\, \citep{Price_MSX_2001}, and has helped in uncovering
its MIR population, especially the deeply embedded stellar objects
\citep{Lumsden_MSX_2002}.

Besides the above,
H$\alpha$ image from the IPHAS\footnote{https://www.iphas.org/images/}
\citep{Drew_IPHAS_2005MNRAS,Barentsen_IPHAS_2014}
survey (Obs date:2003-08-14);
the DSS2-Red image\footnote{https://archive.eso.org/dss/dss};
and the NVSS \citep[NRAO VLA Sky Survey;][]{Condon_NVSS} 1.4\,GHz continuum
image was downloaded for a visual examination of the region.

\section{Results}
\label{section_results}

\subsection{Channel Maps}
\label{section_chmaps}

\begin{figure*}
\includegraphics[width=\textwidth]{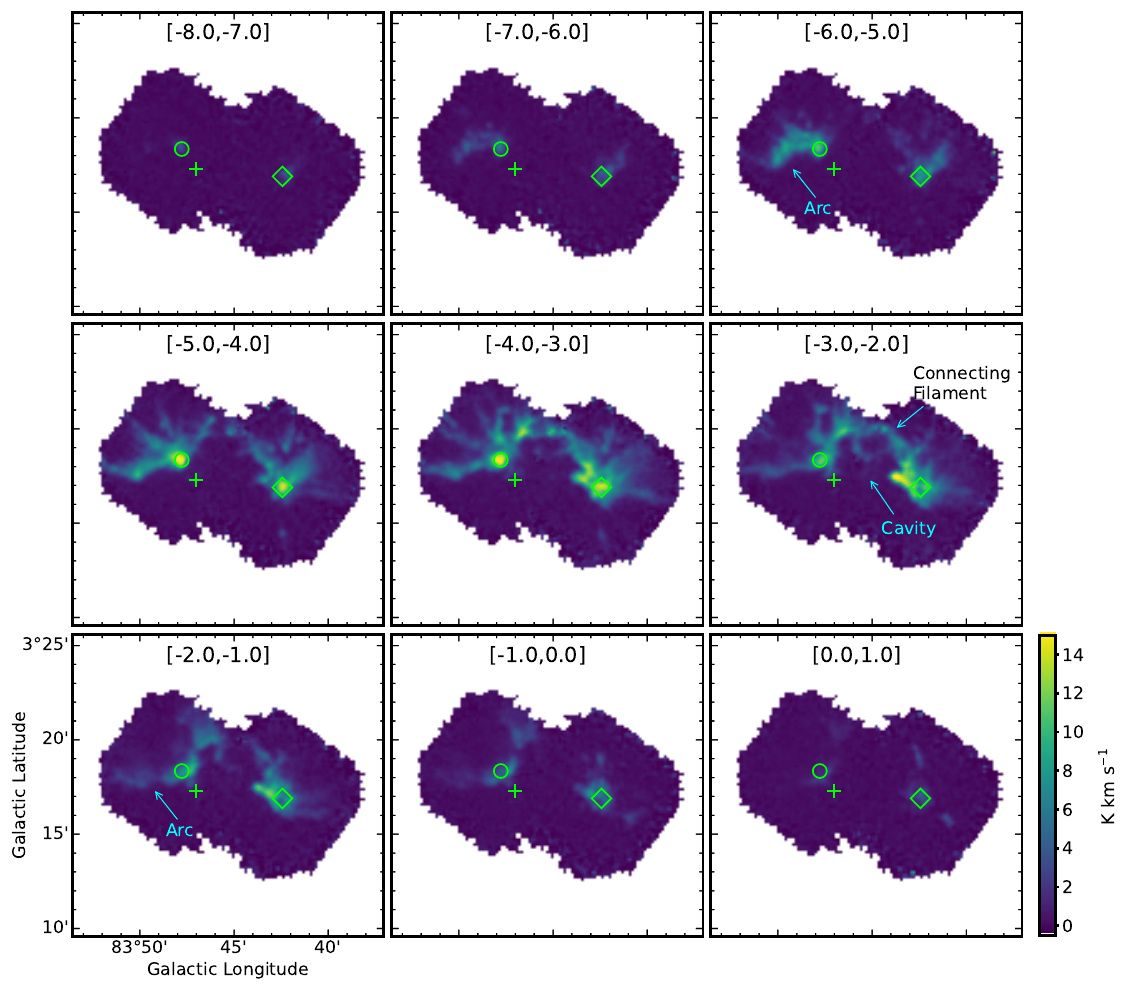}
\caption{Channel maps of \tco\, emission. Plus symbol marks the massive
star \bdstar. Circle and diamond denote the locations of the two RMS
sources, \rmsHII\, and \rmsYSO, respectively.
An arc like filamentary structure, a cavity like structure, and a
connecting filament have been indicated by arrows.
}
\label{fig_chmap_mosaic}
\end{figure*}

Figure \ref{fig_chmap_mosaic} shows the channel map for the Sh2-112
region in \tco\, emission.
The massive star \bdstar, and the two RMS sources
(\rmsHII\, and \rmsYSO) have been marked on the image.
The filamentary nature of the molecular emission is clearly brought
out in the channel maps. While in the eastern part, different
filamentary structures seem to join at one end at the RMS source
\rmsHII\, (circle); in the western portion such structures
seem to converge on \rmsYSO\, (diamond symbol). In the velocity range
-4.0 to -2.0 \kms, the two parts seem to be connected by
molecular filaments. Another noticeable feature is that in the
eastern part, the molecular emission forms an arc like structure
directed away from the massive star \bdstar\,(plus symbol), best
delineated in the [-2.0,-1.0] \kms\, panel. While in the velocity
range [-2.0,0.0]\kms, this arc appears to be directed towards north;
at the blueshifted velocities, for example [-7.0,-5.0]\kms\, range,
it appears curved towards the south. In the middle velocity ranges,
[-5.0,-2.0]\kms, no proper pattern can be made out. It is possible
that two separate filaments, seen at blueshifted and redshifted
velocities, respectively, are merging at this juncture.
Towards the north-west of the massive star \bdstar, there appears
to be a cavity where hardly any molecular emission is seen.
Furthermore, the western emission centered on
the \rmsYSO\, source is also directed away from the massive star.

\subsection{Moment Maps}
\label{section_momentmaps}

In this section, we examine the
m-0 (moment-0 or Integrated Intensity),
m-1 (moment-1 or Intensity-weighted velocity), and
linewidth (Intensity-weighted dispersion) maps
of the molecular emission in the Sh2-112 region.
The examination is confined to only those regions which are
detected at $\geq$\,5$\sigma$ detection level in the spectral
cubes ($\sigma$ being the rms noise level of the respective cubes).
To achieve this, the clumpfind algorithm of
\citet{Williams_1994} was utilised via \textsc{starlink} software's
\citep{Currie_starlink_2014} \textsc{cupid} package \citep{Berry_2007}.
The 5$\sigma$ threshold was conservatively chosen to eliminate
false detections of clumps. As recommended by
\citet{Williams_1994}, the gap between contour levels was kept at
2$\sigma$. With these values, the clumpfind algorithm was
implemented on the mosaiced \co, \tco, and \cetno\,
position-position-velocity cubes. After the detection of clumps,
we mask those regions where no emission was detected at
5$\sigma$ or above to create a masked cube. Thereafter, the respective
(masked) cubes were collapsed to create m-0, m-1, and linewidth
images.

\begin{figure*}
\includegraphics[width=\textwidth]{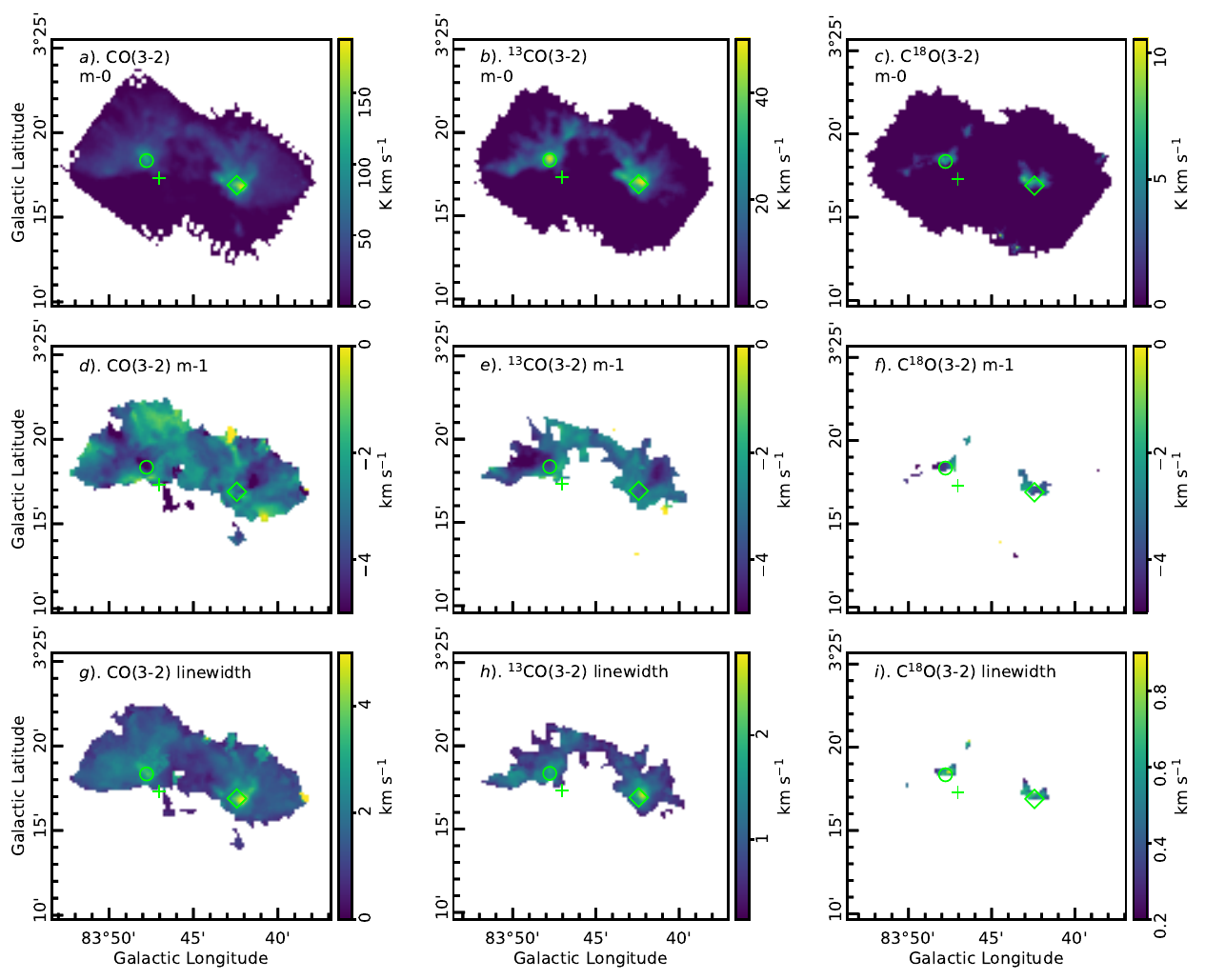}
\caption{Row-wise :
moment-0 (Integrated intensity),
moment-1 (Intensity-weighted velocity), and
linewidth (Intensity-weighted dispersion)
collapsed images for three cubes -- \co, \tco, \cetno\, in first, second, and
third columns, respectively.
The symbols are same as Figure \ref{fig_chmap_mosaic}.
}
\label{fig_m012}
\end{figure*}

The maps are shown in Figure \ref{fig_m012} for \co, \tco, and \cetno\,
molecular transitions. The massive star \bdstar, and the two RMS sources
(\rmsHII\, and \rmsYSO) have been marked on each of the images.
In the m-0 maps, \co\, emission appears to be diffused with no
particular discernible structure, which is expected given the
ubiquitous nature of CO molecule and comparatively lower critical
density than the other two transitions. However, it is still worth
noting that much of the \co\, emission still ``faces away'' from the
massive star \bdstar. The \tco\, m-0 map most clearly traces the
molecular filamentary structures which were examined in the channel maps
(Figure \ref{fig_chmap_mosaic}). While the cavity to the west of the
massive star shows no \tco\, emission, the connecting filamentary
structure between the eastern and western regions is prominent.
Given that all emission here is at $\geq$ 5$\sigma$ level, this
indicates that there is significant molecular mass concentration
in the connecting filament and it is not some minor
emission structure.
The critical density is highest for \cetno\, transition, and thus
only the densest clumps are detected here. Unsurprisingly, the bulk
of \cetno\, emission is associated with the positions of RMS
sources.
The prominent arc like filamentary structure marked
on the \tco\, channel map (Figure \ref{fig_chmap_mosaic})
is also partly seen in \cetno\, m-0 map, indicating
the presence of dense gas along it.
The linewidth maps for all three transitions show a maxima towards
the positions of the RMS sources as compared to the outer parts,
which could be due to outflow activity and/or possibly suggest a convergence
of different flows towards these sources. In the \tco\, linewidth
map, the connecting filament has a relatively lower velocity dispersion,
while the arc like filamentary structure (Figure \ref{fig_chmap_mosaic})
displays a relatively larger dispersion
along its length as compared to others.

\subsection{Physical Parameters}
\label{section_parameters}

\begin{figure*}
\includegraphics[width=\textwidth]{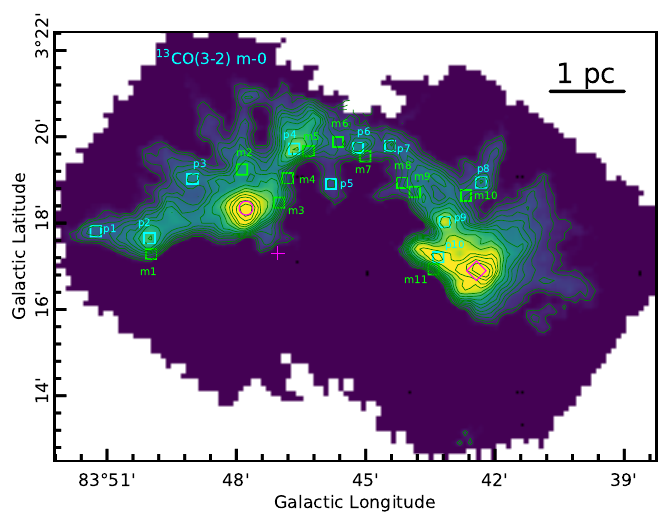}
\caption{
\tco\, integrated intensity map
(contours at 2, 3.5, 6, 8, 10, 12, 14, 17, 21, 25, 30, 34, 40, and 45
K\,km\,s$^{-1}$).
Plus symbol marks the massive star \bdstar.
The two RMS sources, \rmsHII\, and \rmsYSO, have been shown by
circle and diamond symbols, respectively.
The MSX sources have been marked in green boxes and labelled
m1 to m11.
Cyan boxes (labelled p1 to p10) are the locations of local contour
peaks where also we extracted the spectra.
}
\label{fig_spectra_locations}
\end{figure*}

Figure \ref{fig_spectra_locations} shows the \tco\, m-0 map of the region
overlaid with the massive star \bdstar; the two RMS sources (\rmsHII\,
and \rmsYSO); and the MSX sources (m1 to m11) which lay in the regions
of emission $\geq$ 5$\sigma$ level (see section \ref{section_momentmaps}).
The local peaks of m-0 emission, based on visual examination of the
contours have also been marked on the image (p1 to p10).
To obtain the physical parameters for these sources (except \bdstar\,
which did not have any molecular emission associated with it),
we extracted the spectra at their locations (from a 2$\times$2 pixel area).
As high-velocity resolution is required for the calculations, the
non-averaged cubes with the native channel width of $\sim$0.05\kms\,
were utilised for spectrum extraction.

\begin{figure*}
\includegraphics[width=\textwidth]{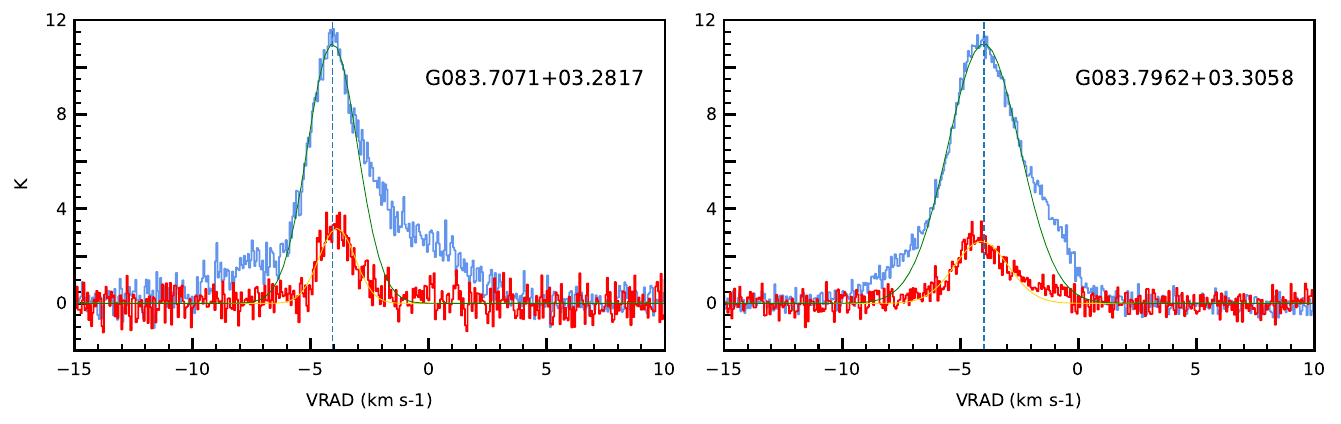}
\caption{
\tco\, (blue) and \cetno\, (red) spectra at positions of
RMS sources marked in Figure \ref{fig_spectra_locations}.
Green and yellow curves show the gaussian fits to the respective
spectra, with blue dashed line marking the velocity of the peak
of the \tco\, gaussian fit.
}
\label{fig_spectra_rms}
\end{figure*}

\begin{figure*}
\includegraphics[width=\textwidth]{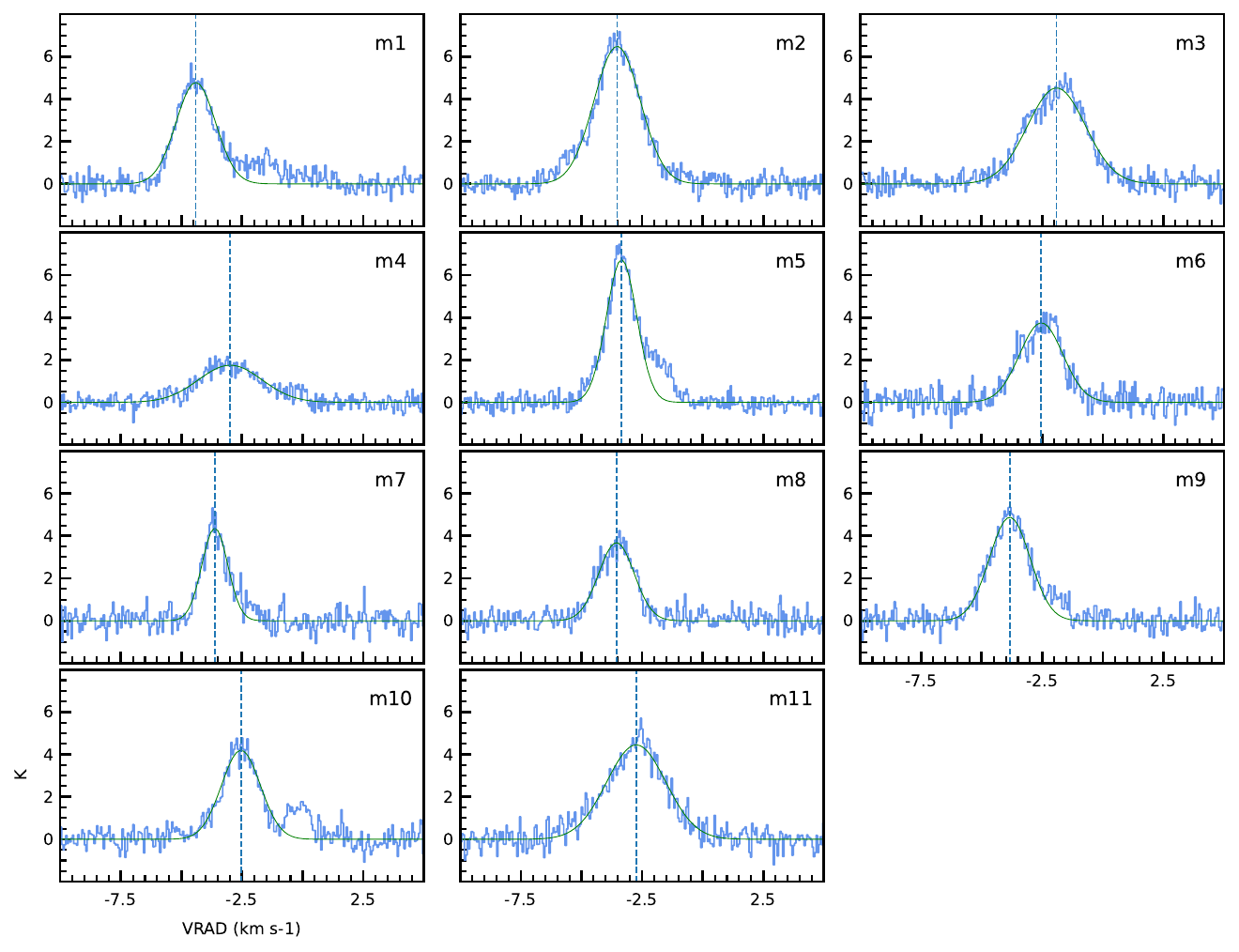}
\caption{
\tco\, spectra at MSX source positions marked in
Figure \ref{fig_spectra_locations} (green boxes labelled m1-m11).
Green curve depicts the gaussian fit to the spectra, with blue dashed line
marking the velocity of the peak of the gaussian fit.
}
\label{fig_spectra_msx}
\end{figure*}

\begin{figure*}
\includegraphics[width=\textwidth]{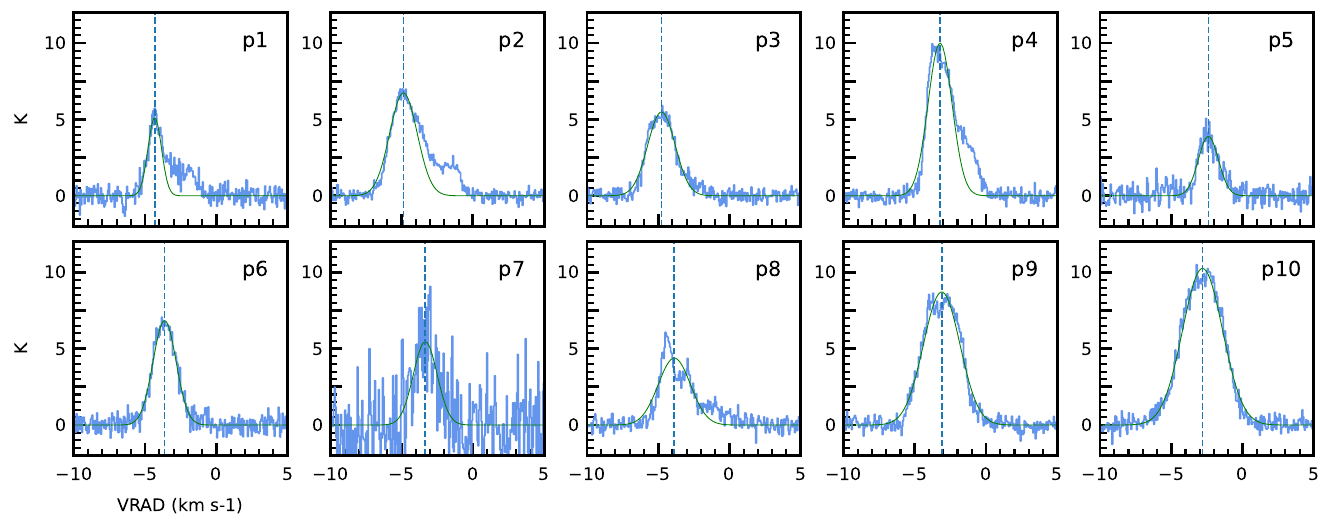}
\caption{
\tco\, spectra at peak positions marked in
Figure \ref{fig_spectra_locations} (cyan boxes labelled p1-p10).
Green curve depicts the gaussian fit to the spectra, with blue dashed line
marking the velocity of the peak of the gaussian fit.
}
\label{fig_spectra_13copeaks}
\end{figure*}

For the RMS sources, both \tco\, and \cetno\, spectra were available,
and are shown in Figure \ref{fig_spectra_rms}. However, for the MSX sources
(m1-m11) and the local \tco\, integrated emission peaks (p1-p10),
only \tco\, spectra could be extracted, shown in
Figures \ref{fig_spectra_msx} and \ref{fig_spectra_13copeaks}, respectively.
The results from the gaussian fitting are given in Table \ref{table_fitting}.
The first thing to notice is that at some of the locations, a simple gaussian
is a poor fit to the spectra. For example, in Figure \ref{fig_spectra_rms},
the spectra have broad wings on both sides, which is usually an indication
of some kind of likely outflow associated with the sources.
According to \citet{Maud_2015MNRAS_Outflows}, while \rmsYSO\, is associated
with an outflow, \rmsHII\, has a possibility of the same.
For some of the locations in
Figures \ref{fig_spectra_msx} and \ref{fig_spectra_13copeaks},
there appears to be excess emission on the red side.
It had been discussed in section \ref{section_chmaps} that the arc like
structure in Figure \ref{fig_chmap_mosaic} probably has a merger of
blueshifted and redshifted filaments. The excess redshifted emission seen
at the locations which lie on this arc,
i.e. m1 (Figure \ref{fig_spectra_msx}), and
p1, p2 (Figure \ref{fig_spectra_13copeaks}) could be an indication of
the same.
Emission peak positions p8 and p9 display significant self-absorption
in their spectra.
The mean velocity for most of the fits lies in the range -2.5 to -4.0\kms.
It is notable that the locations towards the eastern side
(namely p1, p2, p3, and m1) have significantly more blueshifted
mean velocities as opposed to other locations.
The RMS sources, though with wide wings, have nearly same
mean velocities ($\sim$\,-4.0\kms), which is in agreement with the
radial velocity for the Sh2-112 region in
\citet{Blitz_1982ApJS,StarkBrand_1989ApJ,BrandBlitz_1993AA}.
Though our locations display a range of mean velocities, the overall
distribution is consistent with other values cited in literature within
error limits
\citep{Dobashi_1994ApJS,Urquhart_13CO_RMS_2008AA,Lumsden2013_RMSPaper,
Maud_2015MNRAS_RMS,Maud_2015MNRAS_Outflows,Panja_2022}.

Using the FWHM (full width at half maxima) returned by the fit, we
calculate the standard deviation (or observed velocity dispersion), the
non-thermal velocity dispersion, and the total velocity dispersion
using the following set of equations
\citep{FullerAndMyers_1992, FiegeAndPudritz_2000} :

\begin{eqnarray}
\Delta V_{tot}^2 &=& \Delta V_{obs}^2 + 8~~ln\,2~~kT \left( \frac{1}{\bar{m}} - \frac{1}{m_{obs}} \right) \\
\Rightarrow \frac{\Delta V_{tot}^2}{8~~ln\,2} &=& \frac{kT}{\bar{m}} + \left( \frac{\Delta V_{obs}^2}{8~~ln\,2} - \frac{kT}{m_{obs}} \right) \nonumber \\
\Rightarrow \sigma_{tot}^2 &=& c_s^2 + \left( \sigma_{obs}^2 - \sigma_\textsc{t}^2 \right) \\
&=& c_s^2 + \sigma_\textsc{nt}^2 ~~.
\end{eqnarray}

In the above equations,
$\Delta V_{obs}$ is the FWHM of the fit;
$\sigma_{obs}$ ($=\Delta V_{obs}/\sqrt{8ln\,2}~$ for gaussian fits) is the
standard deviation (or dispersion);
$\sigma_\textsc{t} (=\sqrt{kT/m_{obs}})$ is the thermal velocity dispersion;
$m_{obs}$ is the mass of the relevant molecule (29 amu and 30 amu for
$^{13}$CO and C$^{18}$O respectively);
$\sigma_\textsc{nt}$ is the non-thermal velocity dispersion;
$c_s$($=\sqrt{kT/\bar{m}}$) is the speed of sound;
$\bar{m}$ is the average molecular weight of the medium (2.37 amu);
and T is the excitation or gas kinetic temperature.

The FWHM shows a wide range at these locations, ranging from $\sim$\,1.5-3.0
\kms\, for most of the sources, with some of the highest values associated
with locations in the vicinity of the two RMS sources, such as m2, m3, m4,
m11, and p10. This tallies with the linewidth map in Figure
\ref{fig_m012}(h). For the above calculations, we take the excitation
temperature T as 20\,K
\citep[see Figure 15 in][]{Panja_2022}.
Furthermore, we also calculate
Mach number ($= \sigma_\textsc{nt}/c_s$) and the ratio of thermal to
non-thermal pressure ($P_\textsc{tnt} = c_s^2/\sigma_\textsc{nt}^2$)
\citep{Lada_ApJ_2003} for each of the locations. The results of these
two calculations show the presence of supersonic motion, and a dominance
of non-thermal pressure in the cloud. The mach number and P$_\textsc{tnt}$
values are inversely correlated, as expected from their dependence on
$\sigma_\textsc{nt}$. Such values would suggest that the emission mechanism
is likely some supersonic non-thermal phenomena, and could be via
turbulence and magnetic fields
\citep{MyersAndGoodman_MagFields_1988, Crutcher_MagFields_1999},
to suggest one such mechanism.

\begin{table*}
\centering
\caption{Parameters derived from \tco\, spectra at the locations marked in
         Figure \ref{fig_spectra_locations}.
         For \rmsYSO\, and \rmsHII, \cetno\, spectrum
         was also available and thus used for calculation as well.}
\label{table_fitting}
\begin{tabular}{lllllllll}
\hline
clump     &       Mean             &       FWHM             &       Amplitude        &       $\sigma_{NT}$    &       Mach     &       $P_{TNT}$    & \emph{l} & \emph{b}  \\
          &       ($km~s^{-1}$)    &       ($km~s^{-1}$)    &       (K)              &       ($km~s^{-1}$)    &       Number   &                    & (deg)  & (deg)   \\
\hline
\multicolumn{9}{c}{G083.7071+03.2817} \\
\cmidrule(lr){1-9}
~~~C$^{18}$O  &       -3.88            &       1.79             &       3.18             &       0.76             &       2.86         &       0.12  & 83.7071 & 03.2817 \\
~~~$^{13}$CO  &       -4.07            &       2.48             &       11.00            &       1.05             &       3.97         &       0.06  & -       & -       \\
\hline
\multicolumn{9}{c}{G083.7962+03.3058} \\
\cmidrule(lr){1-9}
~~~C$^{18}$O  &       -4.11            &       2.69             &       2.62             &       1.14             &       4.30         &       0.05  & 83.7962 & 3.3058  \\
~~~$^{13}$CO  &       -4.00            &       3.58             &       10.99            &       1.52             &       5.73         &       0.03  &    -    &   -     \\
\hline
\multicolumn{9}{c}{MSX Sources} \\
\hline
m1        &       -4.41            &       1.92             &       4.78             &       0.81             &       3.06         &       0.11  & 83.8330 &  3.2881   \\
m2        &       -3.53            &       2.24             &       6.48             &       0.95             &       3.59         &       0.08  & 83.7978 &  3.3207   \\
m3        &       -1.91            &       2.81             &       4.52             &       1.19             &       4.49         &       0.05  & 83.7835 &  3.3079   \\
m4        &       -2.99            &       3.09             &       1.75             &       1.31             &       4.95         &       0.04  & 83.7803 &  3.3174   \\
m5        &       -3.35            &       1.47             &       6.74             &       0.62             &       2.34         &       0.18  & 83.7721 &  3.3279   \\
m6        &       -2.54            &       2.09             &       3.75             &       0.89             &       3.34         &       0.09  & 83.7607 &  3.3313   \\
m7        &       -3.62            &       1.25             &       4.38             &       0.53             &       1.99         &       0.25  & 83.7502 &  3.3258   \\
m8        &       -3.56            &       1.72             &       3.68             &       0.73             &       2.74         &       0.13  & 83.7359 &  3.3156   \\
m9        &       -3.84            &       1.97             &       4.88             &       0.83             &       3.15         &       0.10  & 83.7310 &  3.3121   \\
m10       &       -2.54            &       1.89             &       4.20             &       0.80             &       3.02         &       0.11  & 83.7112 &  3.3107   \\
m11       &       -2.74            &       2.87             &       4.45             &       1.22             &       4.60         &       0.05  & 83.7237 &  3.2822   \\
\hline
\multicolumn{9}{c}{m-0 contour peaks} \\
\hline
p1        &       -4.32            &       1.11             &       5.14             &       0.47             &       1.76         &       0.32  & 83.8544 &  3.2968   \\
p2        &       -4.87            &       2.28             &       6.74             &       0.96             &       3.64         &       0.08  & 83.8336 &  3.2942   \\
p3        &       -4.77            &       2.27             &       5.49             &       0.96             &       3.62         &       0.08  & 83.8170 &  3.3171   \\
p4        &       -3.22            &       1.97             &       10.01            &       0.83             &       3.15         &       0.10  & 83.7776 &  3.3288   \\
p5        &       -2.39            &       1.67             &       3.89             &       0.70             &       2.66         &       0.14  & 83.7634 &  3.3152   \\
p6        &       -3.63            &       1.91             &       6.85             &       0.81             &       3.05         &       0.11  & 83.7530 &  3.3291   \\
p7        &       -3.34            &       1.97             &       5.46             &       0.83             &       3.14         &       0.10  & 83.7405 &  3.3299   \\
p8        &       -3.87            &       2.66             &       4.38             &       1.12             &       4.25         &       0.05  & 83.7053 &  3.3157   \\
p9        &       -3.11            &       3.01             &       8.72             &       1.28             &       4.82         &       0.04  & 83.7194 &  3.3005   \\
p10       &       -2.80            &       3.30             &       10.26            &       1.40             &       5.29         &       0.04  & 83.7219 &  3.2869   \\
\hline
\end{tabular}
\end{table*}


\section{Discussion}
\label{section_discussion}

\begin{figure*}
\includegraphics[width=\linewidth]{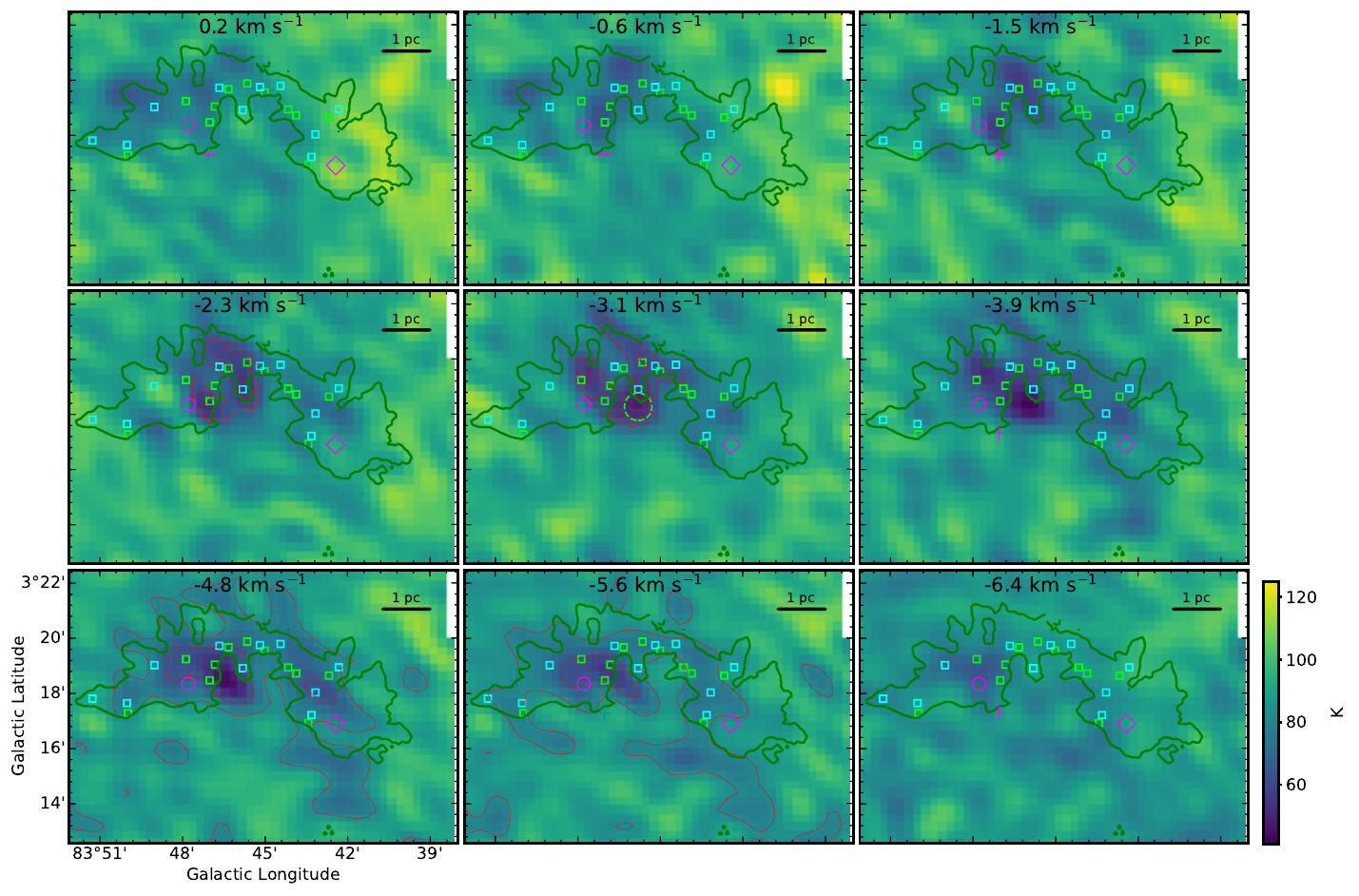}
\caption{
Velocity channel maps for the CGPS H\,I 21\,cm emission.
Green contour marks the 2\,K\,km\,s$^{-1}$ level of the
\tco\, m-0 (Figure \ref{fig_spectra_locations}) image.
In -2.3 and -3.1\kms\, channels, the red contours have been drawn at
(outer to inner)
65, 60, 55, and 50\,K; while in -4.8 and -5.6\kms\, channels, the
red contours are at 81 (outer) and 65 (inner)\,K.
The dashed green circle in -3.1\kms\, channel shows the location where
spectrum was extracted (see Figure \ref{fig_hisa}).
The rest of the symbols are same as Figure \ref{fig_spectra_locations}.
}
\label{fig_cgpsHI}
\end{figure*}

\begin{figure}
\includegraphics[width=0.9\linewidth]{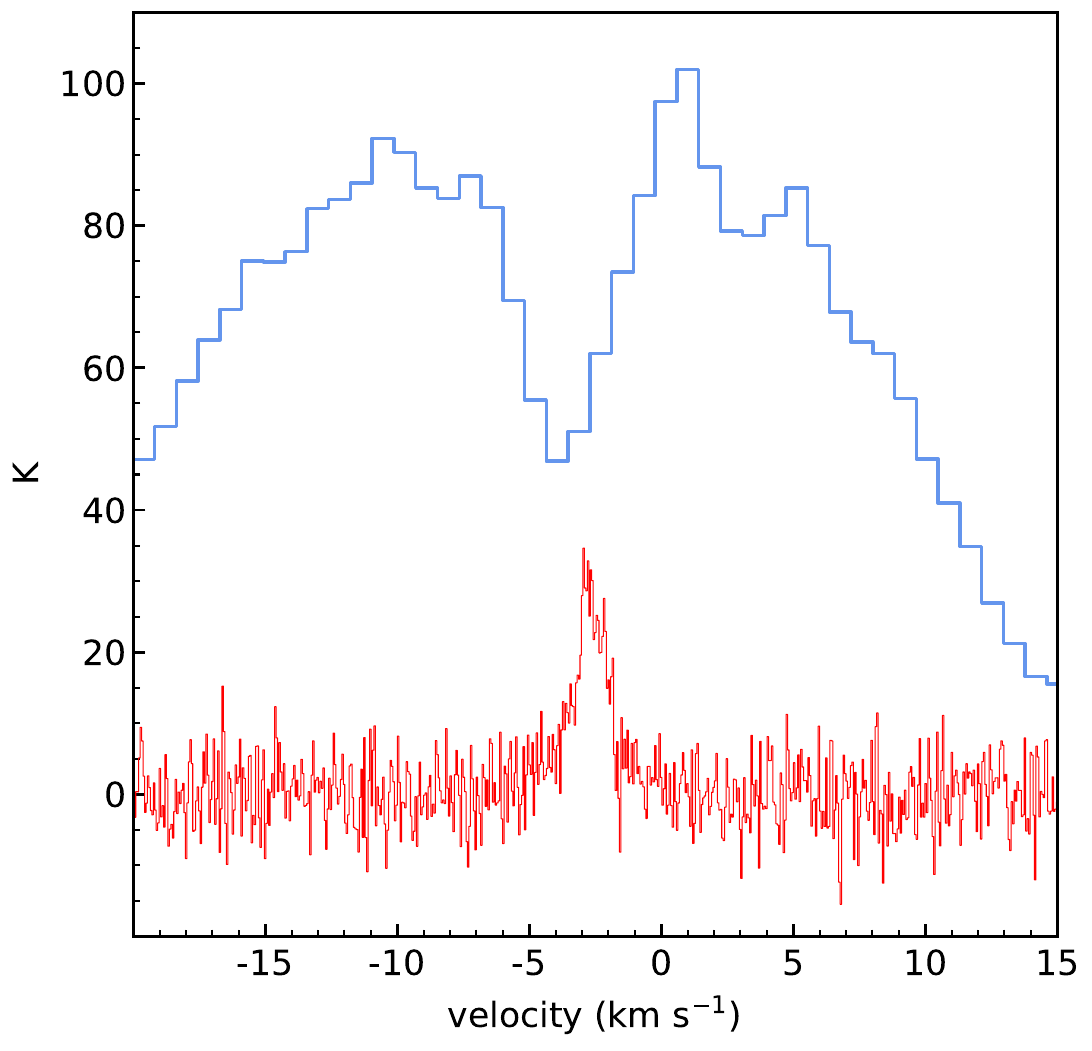}
\caption{
\tco\, spectrum (in red) and
CGPS H\,\textsc{i} spectrum (in blue)
at the location marked in Figure \ref{fig_cgpsHI}
(green circle in -3.1\kms\, channel) -- demonstrating the
H\,\textsc{i} self-absorption feature centered at
$\sim$\,[-5,-4]\kms\, and its anti-correlation
with the molecular emission.
The \tco\, spectrum has been scaled up by a factor of 30$\times$
for better visibility.
}
\label{fig_hisa}
\end{figure}

\begin{figure}
\includegraphics[width=\linewidth]{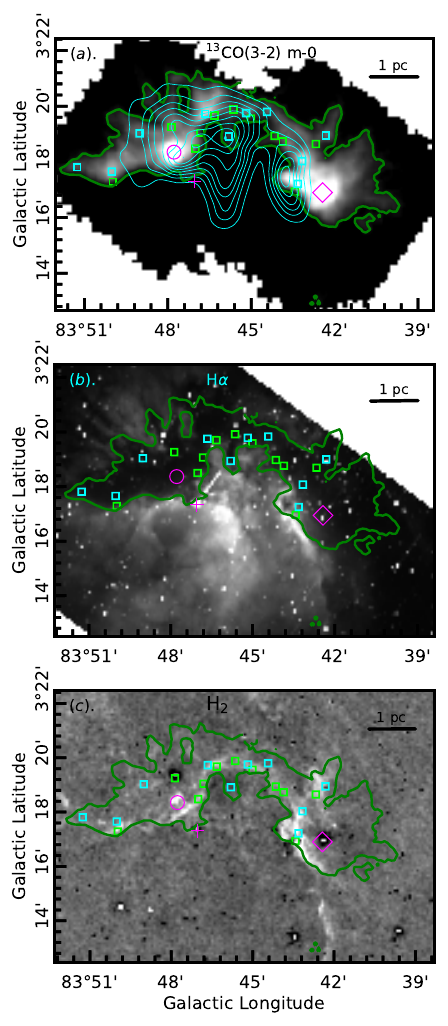}
\caption{
\emph{(a)} \tco\, m-0 map.
Cyan contours are NVSS levels at 0.01, 0.02, 0.03, 0.04, 0.05, 0.06,
0.07, 0.08, 0.09, and 0.1 Jy/beam.
\emph{(b)} H$\alpha$ image of the region.
\emph{(c)} Continuum-subtracted H$_2$ emission (2.12\,$\mu$m) map
from \citet{Panwar_2020}.
Green contour marks the 2\,K\,km\,s$^{-1}$ level of the
\tco\, m-0 (Figure \ref{fig_spectra_locations}) image.
The rest of the symbols are same as Figure \ref{fig_spectra_locations}.
}
\label{fig_COm0_Halpha}
\end{figure}

\begin{figure*}
\includegraphics[width=\linewidth]{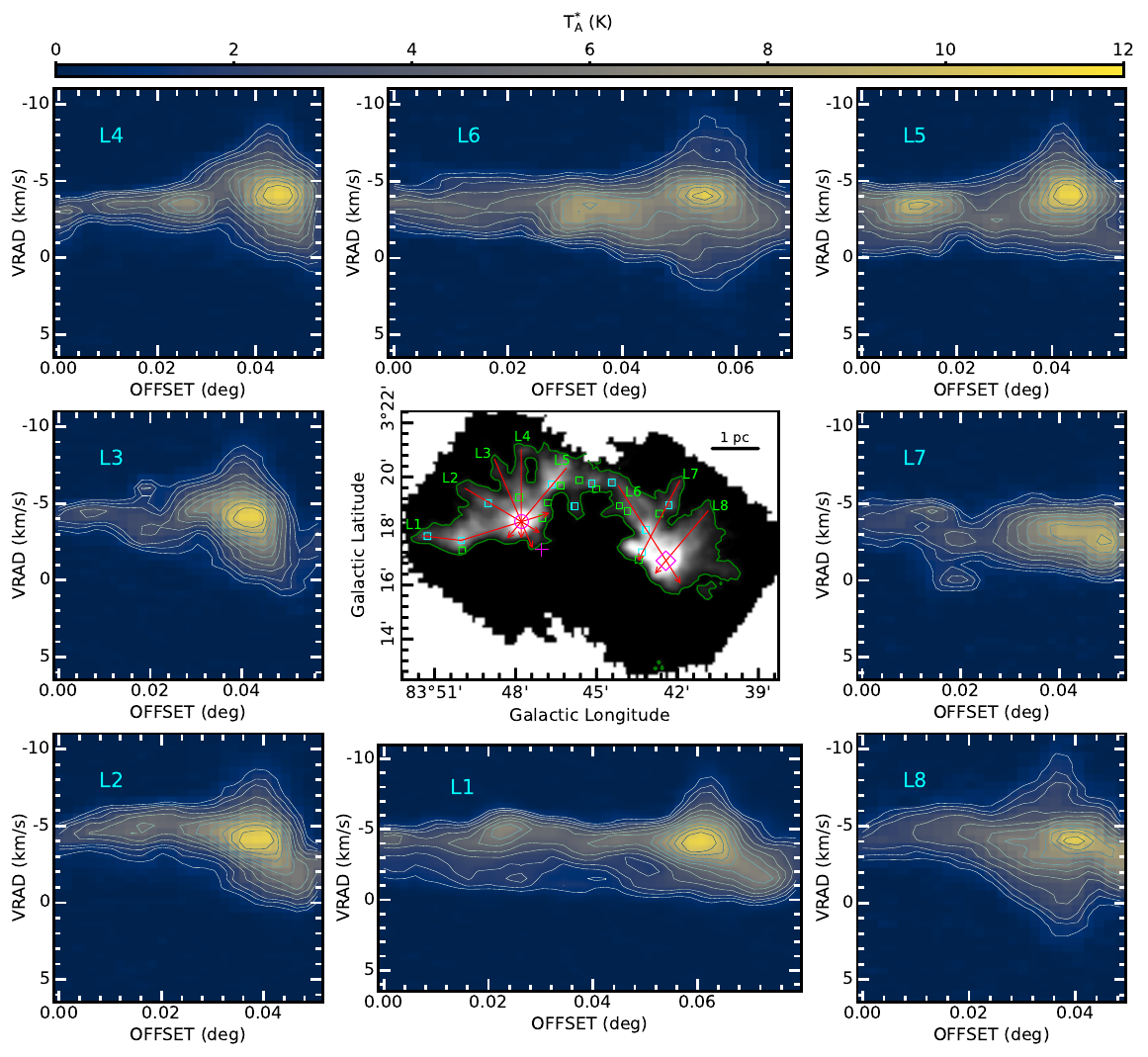}
\caption{Position-velocity diagrams for the line segments
marked L1-L8 (in red) on the central image
(\tco\, m-0 from Figure \ref{fig_spectra_locations}).
Green contour marks the 2\,K\,km\,s$^{-1}$ level of the
\tco\, m-0 (Figure \ref{fig_spectra_locations}) image.
The rest of the symbols are same as Figure \ref{fig_spectra_locations}.
On the L1-L8 p-v maps, the contour levels are at
1, 1.5, 2, 3, 4, 5, 6, 7, 8, 9, and 10\,K.
}
\label{fig_pv}
\end{figure*}

Figure \ref{fig_cgpsHI} shows the CGPS 21\,cm velocity channel maps of this
region. The channel maps show a depression in H\,\textsc{i} emission
(see Figure \ref{fig_hisa}),
which is
(anti-)correlated with strong (i.e.\,$\geq$\,5$\sigma$) molecular emission.
Such features have been referred to as H\,\textsc{i} self absorption
(or HISA) features in literature, and have been found to be an extensive
presence in 21\,cm surveys
\citep{Kerton_HISA_2005ApJ,Wang_HISA_2020AA}.
HISA regions indicate the presence of cold H\,I gas in foreground against
warm H\,I emission from background.
In our channel maps, most intense depression in emission seems to be
confined to the north-eastern quadrant, in the vicinity of the source
\rmsHII\, (circle). Channels -2.3\kms\, and -3.1\kms\, display shell-like
feature on a smaller spatial scale, while the channels -4.8\kms\, and
-5.6\kms\, show large scale shell-like feature. The anti-correlation of
molecular emission and depression in H\,I emission can be most prominently
seen in the -4.8\kms\, and -5.6\kms\, channels. Such shell-like features
have been found in other regions as well, such as
Sh2-237 \citep{Dewangan_S237_2017ApJ},
W4 and W5 \citep{Hosokawa_W4W5_2007ApJ}, and
in Southern Galactic Plane Survey (SGPS) regions \citep{McClure_HISA_2001PASA}.

Figure \ref{fig_COm0_Halpha} shows
the \tco\, m-0 image (overlaid with NVSS contours),
the H$\alpha$ image of the region, and
the continuum-subtracted H$_2$ emission map of the region from
\citet{Panwar_2020}.
The radio emission seems to have multiple peaks, and on a comparison of
Figures \ref{fig_spectra_locations} and \ref{fig_COm0_Halpha}(a),
the radio peaks appear to be associated with the locations
of m3/m4, p5, and a location to the east of p10. According to
\citet{Panwar_2020}, the location to the east of p10 is the ionized
boundary layer.
Based on the presence of this ionized boundary layer and pressure calculations,
they further conjecture the possibility of triggered star formation towards
the location of RMS source \rmsYSO\, due to O8V type star \bdstar\,
\citep[also see][]{Morgan_2004AA,Morgan_2009MNRAS,Urquhart_2009AA}.
It is possible that the expansion of the ionized gas could have had a role
in the formation of the cavity towards the west/northwest of the massive
star \bdstar.
There is significant H$\alpha$ emission in the
cavity, while the H$_2$ 2.12$\mu$m emission seems to trace the southern
boundary of the molecular cloud.
All along the cavity perimeter, there are finger-like filamentary structures
protruding into it, reminiscent of bright-rimmed clouds which are seen in
optical and infrared emission
\citep{Chauhan_2011MNRAS,SaurabhSharma_BRC_2016AJ}.
Here these structures
are seen in absorption in optical (Figure \ref{fig_COm0_Halpha}(b)) and
in emission in \tco\, integrated intensity map
(Figure \ref{fig_COm0_Halpha}(a)).
Two out of three of these finger-like structures seem to be hosting
molecular peaks at their ends, i.e. p5 and p10. If there is a possible
case of triggering towards the source \rmsYSO,
then based on size considerations of the cavity, there could be possible
triggering all along the cavity perimeter. The MSX sources m3, m4, m5, m7,
m8, m9, and m11 seem to be distributed along the cavity boundary where
there is seemingly compression of molecular material, as evidenced by the
sharp change in \tco\, contour levels (see Figure \ref{fig_spectra_locations}).

Lastly we present the position-velocity maps in Figure \ref{fig_pv} along
the different line segments marked L1-L8 on the central image.
While the lines L1-L5 trace the filamentary structures which converge
on the source \rmsHII\, (circle), L6-L8 are cuts along the filamentary
structures associated with \rmsYSO\, (diamond symbol).
The position-velocity maps show a complex structure with clumping along
their length. Significant velocity gradients can also be seen in most of
them, and especially in L1-L5 in the vicinity of the source
\rmsHII\, (i.e. bright clump towards the right in L1-L5).
Here we note that such a gradient could indicate gas being channeled
towards the source, and such a flow of gas along the filamentary structures
towards the RMS sources -- \rmsHII\, and \rmsYSO\, -- as also discussed
in section \ref{section_momentmaps}, is similar to the longitudinal flow in
filaments towards hubs in hub-filament systems
\citep{Dewangan_S237_2017ApJ,Dewangan_G1888_2020ApJ,Williams_SDC13_2018AA}.
Therefore, this region could present a good example of exploring various
star formation frameworks such as
global hierarchical collapse \citep[GHC][]{VazquezSemadeni_GHC_MNRAS_2019},
conveyor belt model
\citep{Longmore_ConveyorBelt_2014, Krumholz_ConveyorBelt_MNRAS_2020},
filaments to clusters model \citep{Nanda_2020}, to name a few.

According to \citet{Panja_2022}, the column density of the molecular
emission region has been calculated to be $\sim$10$^{22}$\,cm$^{-2}$.
Such regions have been designated as ``hubs'' in the context of hub-filament
systems in literature \citep{Myers_2009ApJ}. The column density maps by
\citet{Panja_2022} also suggest such a hub-filament configuration for the
larger region of the order of a few 10s of parsec encompassing Sh2-112.
Thus, what we have explored here seems to be the detailed structure of
a hub region where massive star formation is going on. However, given the
importance of filamentary structures in star formation and expositions
of various filament types in literature
\citep[see][for a detailed review]{Hacar_PP7_2022}, it is essential to
further study such regions via high spectral and spatial resolution
molecular transitions of other species, magnetic fields, and so on so as
to understand the evolution of molecular clouds as they form stars.


\section{Summary and Conclusions}
\label{section_summary}

We have carried out an analysis of the Sh2-112 region in \co, \tco, and
\cetno\, molecular line transitions from the JCMT, supported by archival
data from CGPS H\,I line, MSX, NVSS, and IPHAS H$\alpha$ for visual
examination.
Our main conclusions are as follows :
\begin{enumerate}

\item
The molecular emission appears filamentary in channel maps and seems to be
directed away from the massive star \bdstar, which
is also located at the edge of
a cavity-like structure.

\item
Multiple local peaks were found associated with the molecular emission in
the \tco\, integrated intensity emission map (m-0) which was generated using
clumps detected above 5$\sigma$ level in the position-position-velocity space.
The linewidth map shows high dispersion associated with the positions of
the RMS sources (\rmsHII\, and \rmsYSO).

\item
Analysis of CGPS 21\,cm H\,\textsc{i} line emission reveals the presence of
shell-like HISA feature, where the molecular emission is nearly
coincident with the depression in H\,\textsc{i} emission.

\item
\tco\, spectra was extracted at the locations of RMS sources, MSX sources,
and the local peaks of emission.
For the RMS sources, \cetno\, spectra was also extracted.
Spectral profile fitting suggests significant deviation from a
gaussian profile for many sources.
All the locations were found to have significant non-thermal
dispersions;
large mach numbers ($\sim$\,2--6) indicating dominance of supersonic
motions within the clumps; and a small thermal to non-thermal pressure
ratio ($\sim$\,0.03--0.3).

\end{enumerate}


\section*{Acknowledgements}

We thank the anonymous referee for a critical reading of the manuscript
and for the suggestions for the improvement of this paper.
DKO acknowledges the support of the Department of Atomic Energy, Government
of India, under project Identification No. RTI 4002.
The James Clerk Maxwell Telescope has historically been operated by the
Joint Astronomy Centre on behalf of the Science and Technology Facilities
Council of the United Kingdom, the National Research Council of Canada
and the Netherlands Organisation for Scientific Research.
This research has made use of the NASA/IPAC Infrared Science Archive,
which is funded by the National Aeronautics and Space Administration and
operated by the California Institute of Technology.
This research made use of data products from the Midcourse Space Experiment.
Processing of the data was funded by the Ballistic Missile Defense Organization
with additional support from NASA Office of Space Science. This research has
also made use of the NASA/ IPAC Infrared Science Archive, which is operated
by the Jet Propulsion Laboratory, California Institute of Technology, under
contract with the National Aeronautics and Space Administration.
This research has made use of the services of the ESO Science Archive Facility.

\vspace{-1em}


\end{document}